\documentclass[aps,pre,11pt,superscriptaddress,showpacs]{revtex4}

\usepackage{amsmath}
\usepackage{graphicx}

\begin{document}

\title{Instability and dynamics of two nonlinearly coupled laser beams in a plasma}

\author{P. K. Shukla}
\affiliation{Centre for Nonlinear Physics, Department of Physics,
Ume\aa~University, SE-90187 Ume\aa, Sweden} \affiliation{Institut
f\"ur Theoretische Physik IV and Centre for Plasma Science and
Astrophysics, Fakult\"at f\"ur Physik und Astronomie,
Ruhr--Universit\"at Bochum, D-44780 Bochum, Germany}

\author{B. Eliasson}\affiliation{Centre for Nonlinear Physics, Department of Physics,
Ume\aa~University, SE-90187 Ume\aa, Sweden}
\affiliation{Institut f\"ur Theoretische Physik IV and Centre for
Plasma Science and Astrophysics, Fakult\"at f\"ur Physik und
Astronomie, Ruhr--Universit\"at Bochum, D-44780 Bochum, Germany}

\author{M. Marklund}
\affiliation{Centre for Nonlinear Physics, Department of Physics,
Ume\aa~University, SE-90187 Ume\aa, Sweden} \affiliation{Institut
f\"ur Theoretische Physik IV and Centre for Plasma Science and
Astrophysics, Fakult\"at f\"ur Physik und Astronomie,
Ruhr--Universit\"at Bochum, D-44780 Bochum, Germany}

\author{L. Stenflo}
\affiliation{Centre for Nonlinear Physics, Department of Physics,
Ume\aa~University, SE-90187 Ume\aa, Sweden}

\author{I. Kourakis}
\affiliation{Institut f\"ur Theoretische Physik IV and Centre for
Plasma Science and Astrophysics, Fakult\"at f\"ur Physik und
Astronomie, Ruhr--Universit\"at Bochum, D-44780 Bochum, Germany}

\author{M. Parviainen}
\affiliation{Institut f\"ur Theoretische Physik IV and Centre for
Plasma Science and Astrophysics, Fakult\"at f\"ur Physik und
Astronomie, Ruhr--Universit\"at Bochum, D-44780 Bochum, Germany}

\author{M. E. Dieckmann}
\affiliation{Institut f\"ur Theoretische Physik IV and Centre for
Plasma Science and Astrophysics, Fakult\"at f\"ur Physik und
Astronomie, Ruhr--Universit\"at Bochum, D-44780 Bochum, Germany}

\received{16 March 2005}

\begin{abstract}
We investigate the nonlinear interaction between two laser beams in
a plasma in the weakly nonlinear and relativistic regime. The
evolution of the laser beams is governed by two nonlinear
Schr\"odinger equations that are coupled with the slow plasma density response.
We study the growth rates of the Raman forward and backward
scattering instabilities as well of the Brillouin and
self-focusing/modulational instabilities. The nonlinear evolution
of the instabilities is investigated by means of direct simulations
of the time-dependent system of nonlinear equations.
\end{abstract}


\maketitle

\section{Introduction}
The interaction between intense laser beams and plasmas
leads to a variety of different instabilities, including
Brillouin and Raman forward and backward 
\cite{Shukla86,Sjolund67,Yu74,Shukla75,Shukla84,Tsintsadze74} scattering and
modulational instabilities. 
In multiple dimensions we also have filamentation and side-scattering 
instabilities. Relativistic effects
can then play an important role \cite{Shukla86,Tsintsadze74,Max74}. 
When two laser
beams interact in the plasma, we have a new set of
phenomena. An interesting application is the beat-wave accelerator, in
which two crossing beams with somewhat different frequencies 
can accelerate electrons to ultra-relativistic speeds via the
ponderomotive force acting on the electrons. The modulational
and filamentation instabilities of multiple co-propagating electromagnetic 
waves can be described by a system of coupled nonlinear Schr\"odinger equations from which the
nonlinear wave coupling and the interaction between localized light wave packets 
can be easily studied \cite{Shukla92,Berge98}. Two co-propagating narrow 
laser beams may attract each other and spiral around each other \cite{Ren01} or merge \cite{Dong02}. 
Counter-propagating laser beams
detuned by twice the plasma frequency can, at relativistic intensities,
give rise to fast plasma waves via higher-order nonlinearities 
\cite{Rosenbluth72,Shvets01,Bingham04}. At relativistic amplitudes, plasma
waves can also be excited via beat wave excitation at frequencies
different from the electron plasma frequency, with applications
to efficient wake-field accelerators \cite{Shvets04}. 
The relativistic wakefield behind intense laser pulses is
periodic in one-dimension \cite{Berezhiani90} and shows a
quasi-periodic behavior in multi-dimensional simulations \cite{Tsung04}.
Particle-in-cell
simulations have demonstrated the generation of large-amplitude 
plasma wake-fields by colliding laser pulses \cite{Nagashima01} 
or by two co-propagating pulses where a long trailing pulse is modulated
efficiently by the periodic plasma wake behind the 
first short pulse \cite{Sheng02}.

In the present paper, we consider the nonlinear
interaction between two weakly relativistic crossing laser
beams in plasmas. We derive a set of nonlinear mode coupled equations and nonlinear dispersion
relations, which we analyze for Raman backward and forward scattering instabilities
as well as for Brillouin and modulation/self-focusing instabilities.

\section{Nonlinear model equations}

We consider the propagation of intense laser light in an electron--ion
plasma.  The slowly varying electron density perturbation is denoted
by $n_{es1}$. Thus, our starting point is the Maxwell equation
\begin{equation}\label{eq:evol}
  \nabla\times\vec{B} = -\frac{4\pi}{c}(n_0 + n_{es1})e\vec{v} + \frac{1}{c}\frac{\partial\vec{E}}{\partial t} .
\end{equation}
The laser field is given in the radiation gauge,
$\vec{B} = \nabla\times\vec{A}$ and $\vec{E} = -({1}/{c}){\partial\vec{A}}/{\partial t}$.
Since $\partial\vec{p}_e/\partial t = -e\vec{E}$, we thus have 
$\vec{p}_e = e\vec{A}/c$.
Moreover, $\vec{p}_e = m_e\gamma\vec{v}_e$, where $m_e$ is the
electron rest mass and $\gamma = (1 - v_e^2/c^2)^{-1/2}$ is the
relativistic gamma factor, so that
\begin{equation}\label{eq:e-vel}
  \vec{v}_e = \frac{e\vec{A}}{m_ec}\left(1 + \frac{2 e^2|\vec{A}|^2}{m_e^2c^4}\right)^{-1/2}.
\end{equation}
For weakly relativistic particles, i.e.\ $e^2|\vec{A}|^2/m_e^2c^4
\ll 1$, we can approximate (\ref{eq:e-vel}) by
\begin{equation}
  \vec{v}_e \approx \frac{e\vec{A}}{m_ec}\left( 1 - \frac{e^2|\vec{A}|^2}{m_e^2c^4} \right) .
\end{equation}
With these prerequisites, Eq.\ (\ref{eq:evol}) becomes
\begin{equation}\label{eq:nlin}
  \left( \frac{\partial^2}{\partial t^2} - c^2\nabla^2 \right) \vec{A}
  + \omega_{\mathrm{p}0}^2\left(1 + N_s \right)\vec{A}  - 
  \omega_{\mathrm{p}0}^2\frac{e^2|\vec{A}|^2}{m_e^2c^4}\vec{A} = 0 ,
\end{equation}
where $\omega_{\mathrm{p}0}=(4\pi n_0 e^2/m_e)^{1/2}$ is the electron plasma frequency and we
have denoted $N_s=n_{es1}/n_0$.

Next, we divide the vector potential into two parts according
to $\vec{A} = \vec{A}_1 + \vec{A}_2$, representing the two laser
pulses. We also consider the case $\vec{A}_1\cdot\vec{A}_2 \approx 0$.
With this, we obtain from (\ref{eq:nlin}) the two coupled equations
\begin{subequations}
\begin{equation}
  \left( \frac{\partial^2}{\partial t^2} - c^2\nabla^2 \right) \vec{A}_1
  + \omega_{\mathrm{p}0}^2\left(1 + N_s \right)\vec{A}_1  - 
  \omega_{\mathrm{p}0}^2\frac{e^2}{m_e^2c^4}(|\vec{A}_1|^2 + |\vec{A}_2|^2)\vec{A}_1 = 0 ,
\end{equation}
and
\begin{equation}
  \left( \frac{\partial^2}{\partial t^2} - c^2\nabla^2 \right) \vec{A}_2
  + \omega_{\mathrm{p}0}^2\left(1 + N_s \right)\vec{A}_2  - 
  \\
  \omega_{\mathrm{p}0}^2\frac{e^2}{m_e^2c^4}(|\vec{A}_1|^2 + |\vec{A}_2|^2)\vec{A}_2 = 0.
\end{equation}
\label{eq:nlin-coupled}
\end{subequations}
Assuming that $A_j$ is proportional to $\exp(i\vec{k}_j\cdot\vec{r}-i\omega_j t)$ where
$\omega_j\gg|\partial/\partial t|$, we obtain in the slowly varying envelope approximation
two coupled nonlinear Schr\"odinger equations
\begin{subequations}
\begin{equation}
  -2i\omega_1 \left( \frac{\partial}{\partial t}+\vec{v}_{g1}\cdot\nabla\right)\vec{A}_1 - c^2\nabla^2 \vec{A}_1
  + \omega_{\mathrm{p}0}^2 N_s \vec{A}_1  - 
  \omega_{\mathrm{p}0}^2\frac{e^2}{m_e^2c^4}(|\vec{A}_1|^2 + |\vec{A}_2|^2)\vec{A}_1 = 0 ,
\end{equation}
and
\begin{equation}
  -2i\omega_2 \left( \frac{\partial}{\partial t}+\vec{v}_{g2}\cdot\nabla\right)\vec{A}_2 - c^2\nabla^2 \vec{A}_2
  + \omega_{\mathrm{p}0}^2 N_s \vec{A}_2  - 
  \\
  \omega_{\mathrm{p}0}^2\frac{e^2}{m_e^2c^4}(|\vec{A}_1|^2 + |\vec{A}_2|^2)\vec{A}_2 = 0 ,
\end{equation}
\label{eq:nlse-coupled}
\end{subequations}
where $\vec{v}_{gj}=\vec{k}_jc^2/\omega_j$ is the group velocity and
$\omega_j=(\omega_{p0}^2+c^2k_j^2)^{1/2}$ is the electromagnetic wave frequency.

In order to close (\ref{eq:nlse-coupled}), we next consider
the slow plasma response. Here we may follow two routes.
First, if we assume immobile ions, the slowly varying electron
number density and velocity perturbations satisfy the equations
\begin{equation}
  \frac{\partial n_{es1}}{\partial t} + n_0\vec{\nabla}\cdot\vec{v}_{es1} =0 ,
  \label{eq7}
\end{equation}
and
\begin{equation}
  \frac{\partial\vec{v}_{es1}}{\partial t} + \frac{e^2}{m_e^2c^2}\vec{\nabla}(|\vec{A}_1|^2+|\vec{A}_2|^2) =
    \frac{e}{m_e}\vec{\nabla}\phi_s -
    \frac{3T_e}{m_en_0}\vec{\nabla}n_{es1}
\end{equation}
where $T_e$ is the electron temperature, together with the Poisson equation
\begin{equation}
  \nabla^2\phi_s = 4\pi en_{es1} .
  \label{eq9}
\end{equation}
Thus, combining (\ref{eq7})--(\ref{eq9}) together with the vector potential
decomposition, we obtain
\begin{equation}\label{eq:acoustic1}
  \left( \frac{\partial^2}{\partial t^2} - 3v_{Te}^2\nabla^2 + \omega_{\mathrm{p}0}^2
  \right)N_s
  = \frac{e^2}{m_e^2c^2}\nabla^2(|\vec{A}_1|^2 + |\vec{A}_2|^2) ,
\end{equation}
where the electron thermal velocity is denoted by $v_{Te} = (T_e/m_e)^{1/2}$.

Second, if the electrons are treated as inertialess, we have in the
quasi-neutral limit $n_{is1}=n_{es1}\equiv n_{s1}$
\begin{equation}\label{eq:inless1}
  \frac{n_0e^2}{m_ec^2}\vec{\nabla}(|\vec{A}_1|^2+|\vec{A}_2|^2) = n_0e\vec{\nabla}\phi_s - T_e\vec{\nabla}n_{s1} ,
\end{equation}
and
\begin{equation}\label{eq:inless2}
  n_0 m_i\frac{\partial\vec{v}_{is1}}{\partial t} = -n_0e\vec{\nabla}\phi_s - 3T_i\vec{\nabla}n_{s1} .
\end{equation}
Adding Eqs.\ (\ref{eq:inless1}) and (\ref{eq:inless2}), we obtain
\begin{equation}
  n_0 m_i\frac{\partial\vec{v}_{is1}}{\partial t} 
  + \frac{n_0e^2}{m_ec^2}\vec{\nabla}(|\vec{A}_1|^2+|\vec{A}_2|^2) + (T_e + 3T_i)\vec{\nabla}n_{s1}=0 ,
\end{equation}
which should be combined with
\begin{equation}
  \frac{\partial n_{s1}}{\partial t} + n_0\vec{\nabla}\cdot\vec{v}_{is1} = 0,
\end{equation}
to obtain
\begin{equation}\label{eq:acoustic2}
  \left( \frac{\partial^2}{\partial t^2} 
  - c_s^2\nabla^2\right)N_s = 
  \frac{e^2}{m_em_ic^2}\nabla^2(|\vec{A}_1|^2 + |\vec{A}_2|^2) ,
\end{equation}
where the sound speed is $c_s = \sqrt{(T_e +
3T_i)/m_i}$ and $T_i$ is the ion temperature.

\section{Coupled laser beam amplitude modulation theory}

We shall consider, successively, Eqs. (6a, b) combined with
(\ref{eq:acoustic1}) (Case I: Raman scattering) or with
(\ref{eq:acoustic2}) (Case II: Brillouin scattering).

\subsection{Evolution equations}

Setting $\nabla \rightarrow i K$ and $\partial/\partial t
\rightarrow - i \Omega$ into the equations for the plasma density
responses, we obtain
\begin{equation}
N_s = \alpha_0 \, (|\vec{A}_1|^2 + |\vec{A}_2|^2)  \, ,
 \label{response}
\end{equation}
where,
\begin{subequations}
for Case I:
\begin{equation}
\alpha_0 = \frac{e^2}{m_e^2 c^2} \, \, \frac{K^2}{\Omega^2 -
3 K^2 v_{Te}^2 - \omega_{p0}^2} \, ,
\end{equation}
and for Case II:
\begin{equation}
\alpha_0  = \frac{e^2}{m_e m_i c^2}  \, \,
\frac{K^2}{\Omega^2 - K^2 c_{s}^2} \, .
\end{equation}
\label{alpha0}
\end{subequations}

The expressions (\ref{response}) and (\ref{alpha0}) derived above
provide the slow plasma response for any given pair of fields $\{
\vec{A}_j \}$ ($j=1, 2$). The latter now obey a set of coupled
equations, which are obtained by substituting (\ref{response}) into
(\ref{eq:nlse-coupled}),
\begin{subequations}
\begin{equation}
  2i\omega_1\left( \frac{\partial}{\partial t}
  + \vec{v}_{g1}\cdot\vec{\nabla} \right)\vec{A}_1 + c^2\nabla^2\vec{A}_1 + \omega_{\mathrm{p}0}^2  \, \biggl( \frac{e^2}{m_e^2 c^4}
- \alpha_0 \biggr) \, (|\vec{A}_1|^2 + |\vec{A}_2|^2)\vec{A}_1 = 0
\, ,
\end{equation}
and
\begin{equation}
   2i\omega_2 \left( \frac{\partial}{\partial t}
   + \vec{v}_{g2}\cdot\vec{\nabla} \right)\vec{A}_2 + c^2\nabla^2\vec{A}_2
   + \omega_{\mathrm{p}0}^2  \biggl( \frac{e^2}{m_e^2 c^4}
- \alpha_0 \biggr) \, (|\vec{A}_1|^2 + |\vec{A}_2|^2)\vec{A}_2 = 0
\, ,
\end{equation}
\label{cnlse1}
\end{subequations}
For convenience, Eqs. (\ref{cnlse1}a) and (\ref{cnlse1}b)
are cast into the reduced form as
\begin{subequations}
\begin{equation}
 2i\omega_1 \left(
\frac{\partial}{\partial t} + \vec{v}_{g1}\cdot\vec{\nabla}
\right)\vec{A}_1 + c^2 \, \nabla^2\vec{A}_1 + Q \, (|\vec{A}_1|^2 +
|\vec{A}_2|^2)\vec{A}_1 = 0 \, , \label{red1}
\end{equation}
and
\begin{equation}
2i\omega_2\left(
\frac{\partial}{\partial t} + \vec{v}_{g2}\cdot\vec{\nabla}
\right)\vec{A}_2 + c^2 \, \nabla^2\vec{A}_2 + Q \, (|\vec{A}_1|^2 +
|\vec{A}_2|^2)\vec{A}_2 = 0 \, , \label{red2}
\end{equation}
\label{cnlse}
\end{subequations}
where $\vec{A}_j$ has been normalized by $m_e c^2/e$ and where
the nonlinearity/coupling coefficients are
\begin{subequations}
\begin{equation}
Q= \omega_{p0}^2 \biggl( 1 - \frac{K^2 c^2}{\Omega^2
- 3 K^2 v_{Te}^2 - \omega_{p0}^2}   \biggr) \, ,
\end{equation}
and
\begin{equation}
Q= \omega_{p0}^2 \biggl( 1 - \frac{m_e}{m_i} \,
\frac{K^2 c^2}{\Omega^2 - K^2 c_s^2}   \biggr) \, ,
\end{equation}
\label{Qcoeff}
\end{subequations}
for stimulated Raman (Case I) and Brillouin (Case II) scattering, respectively.
We observe that the expressions (\ref{Qcoeff}a) and (\ref{Qcoeff}b)
may be either positive or negative, depending on the
frequency $\Omega$, prescribing either the modulational instability
or the Raman and Brillouin scattering instabilities \cite{Tsintsadze79}.

The two nonlinear wave equations are identical upon an index ($1,
2$) interchange, and coincide for equal frequencies $\omega_1 =
\omega_2$.

\subsection{Nonlinear dispersion relation}

We now investigate the parametric instabilities of the system of
equations (\ref{red1}) and (\ref{red2}). Fourier decomposing the
system by the ansatz
$\vec{A}_j=[\vec{A}_{j0}+\vec{A}_{j+}\exp(i\vec{K}\cdot\vec{r}-i\Omega
t)+\vec{A}_{j-}\exp(-i\vec{K}\cdot\vec{r}+i\Omega
t))]\exp(-i\Omega_0 t)$, where $|\vec{A}_{j0}|\gg |\vec{A}_{j\pm}|$,
and sorting for different powers of
$\exp(i\vec{K}\cdot\vec{r}-i\Omega t)$, we find the nonlinear
frequency shift
\begin{equation}
  \Omega_{j0}=-Q_{K=0}(|\vec{A}_{10}|^2+|\vec{A}_{20}|^2)/2 \omega_j,
\label{nonlin}
\end{equation}
where $Q_{K=0}$ denotes the expression for $Q$ with $K=0$. For the nonlinear
wave couplings, we have from (\ref{cnlse}) the system of equations
\begin{subequations}
\begin{eqnarray}
&& D_{1+}X_{1+}+Q|A_{10}|^2(X_{1+}+X_{1+}+X_{2+}+X_{2-})=0,
\\
&&D_{1-}X_{1-}+Q|A_{10}|^2(X_{1+}+X_{1+}+X_{2+}+X_{2-})=0,
\\
&&D_{2+}X_{2+}+Q|A_{20}|^2(X_{1+}+X_{1+}+X_{2+}+X_{2-})=0,
\\
&&D_{1-}X_{1-}+Q|A_{20}|^2(X_{1+}+X_{1+}+X_{2+}+X_{2-})=0,
\end{eqnarray}
\label{nonlinsys}
\end{subequations}
where the unknowns are $X_{1+}=\vec{A}_{10}^*\cdot\vec{A}_{1+}$,
$X_{1-}=\vec{A}_{10}\cdot\vec{A}_{1-}^*$,
$X_{2+}=\vec{A}_{20}^*\cdot\vec{A}_{2+}$, and
$X_{2-}=\vec{A}_{20}\cdot\vec{A}_{2-}^*$.
The sidebands are characterized by
\begin{equation}
D_{j\pm}=\pm
2[\omega_j\Omega-c^2\vec{k}_{j}\cdot\vec{K}]-c^2K^2,
\end{equation}
where we have used $\vec{v}_{gj}=c\vec{k}_j/\omega_j$.
The solution of the system of equations (\ref{nonlinsys}) yields the 
nonlinear dispersion relation
\begin{equation}
\frac{1}{Q}+\left(\frac{1}{D_{1+}}+\frac{1}{D_{1-}}\right)|\vec{A}_{10}|^2
+\left(\frac{1}{D_{2+}}+\frac{1}{D_{2-}}\right)|\vec{A}_{20}|^2=0,
\label{nonlindisp}
\end{equation}
which relates the complex-valued frequency $\Omega$ to the
wavenumber $\vec{K}$.
Equation (\ref{nonlindisp}) covers Raman forward and
backscattering instabilities, as well as the Brillouin backscattering
instability or the modulational/self-focusing instability, depending
on the two expressions for the coupling constant $Q$. If either
$|\vec{A}_{10}|$ or $|\vec{A}_{20}|$ is zero, then we recover the
usual expressions for a single laser beam in a laboratory
plasma, or for a high-frequency radio beam in the ionosphere \cite{Stenflo90}.

\section{Numerical results}

We have solved the nonlinear dispersion relation (\ref{nonlindisp}) 
and presented the numerical results in Figs. 1--5.
In all cases, we have used the normalized weakly relativistic pump wave amplitudes $A_{j0}=0.1$
with different sets of wavenumbers for the two beams. 
The nonlinear couplings between the laser beams and the Langmuir waves,
giving rise to the Raman scattering instabilities (Case I), are considered
in Figs. 1 and 2. The instability essentially obeys the matching conditions
$\omega_{j}=\omega_{s}+\Omega$ and 
$\vec{k}_{j}=\vec{k}_{s}+\vec{K}$, where
$\omega_{j}$ and $\vec{k}_{j}$ are the frequency and wavenumbers of the pump wave,
$\omega_{s}$ and $\vec{k}_{s}$ are the frequency and wavenumbers for the 
scattered and frequency downshifted electromagnetic daughter wave,
$\Omega$ and $K$ are the frequencies of the
Langmuir waves, and where the light waves approximately obey the
linear dispersion relation, $\omega_{j}=(\omega_{p0}^2+ k_j^2 c^2)^{1/2}$,
$\omega_{s}=(\omega_{p0}^2+k_{s}^2 c^2)^{1/2}$ and the low-frequency waves
obey the Langmuir dispersion relation $\Omega=(\omega_{p0}^2+3K^2 v_{Te}^2)^{1/2}$. 
We thus have the matching condition 
$(\omega_{p0}^2+k_{j}^2 c^2)^{1/2}=
[\omega_{p0}^2+(\vec{k}_{j}-\vec{K})^2 c^2 ]^{1/2}+(\omega_{p0}^2+3 K^2 v_{Te}^2)^{1/2}$, which
in two-dimensions relates the components $K_y$ and $K_z$ of the Langmuir
waves to each other, and which gives rise to almost circular regions 
of instability, as seen in Figs. 1 and 2. In the upper left and right panels of Fig. 1, we
have assumed that the single beams $A_1$ and $A_2$ propagate in the $y$ and $z$ direction,
respectively, having the wavenumber $(k_{1y},k_{1z})=(6,\,0)$ 
and $(k_{2y},\,k_{2z})=(0,\,4)$, respectively. We can clearly see a
backward Raman instability, which for the beams $A_1$ and $A_2$ have maximum
growth rates at $(K_y,K_z)=(2 k_{1y},\,0)=(12,\,0)\,\omega_{p0}/c$ and 
$(K_y,\,K_z)=(0,\,2k_{2z})=(0,\,8)\,\omega_{p0}/c$, respectively.
The backward Raman instability is connected via the obliquely growing wave modes
to the forward Raman scattering instability that has a maximum growth rate (much smaller than that
of the backward Raman scattering instability) at the wave number
$K\approx \omega_{pe}/c$ in the same directions as the laser beams. 
In the lower panels, we consider the two beams propagating simultaneously in the plasma,
at a right angle to each other (lower left panel) and in opposite directions 
(lower right panel). We see that the dispersion relation predicts a rather weak interaction
between the two laser beams, where the lower left panel shows more or less a
superposition of the growth rates in the two upper panels. The case of two 
counter-propagating laser beams (lower right panel) also shows a weak interaction between the two beams.
For the case of equal wavelengths of the two pump waves, as shown in Fig. 2, we have a
similar scenario as in Fig. 1. The lower left panel of Fig. 2 shows that the
growth rate of two interacting laser beams propagating at a right angle to each other
is almost a superposition of the growth rates of the single laser beams displayed
in the upper panels of Fig. 2. Only for the counter-propagating laser beams in the
lower right panel we see that the instability regions have split into broader and narrower
bands of instability, while the magnitude of the instability is the same as for the single beam cases.

We next turn to the Brillouin scattering scenario (Case II), in which 
the laser wave is scattered against ion acoustic waves, displayed in Figs. 3 and 4.
In the weakly nonlinear case, we have three-wave couplings in the same manner
as for the interaction with Langmuir waves, and we see in both Figs. 3 and 4
that the instability has a maximum growth rate in a narrow, almost circular band
in the $(K_y,K_z)$ plane. In the upper two panels, we also see the backscattered Brillouin instability
with a maximum growth rate at approximately twice the pump wavenumbers, but we do not
have the forward scattered instability. Instead, we see a broadband weak instability
in all directions and also perpendicular to the pump wavenumbers. A careful study
shows that the perpendicular waves are purely growing, i.e. there may be density channels
created along the propagation direction of the laser beam. In the lower panels of 
Figs. 3 and 4, we display the cases with interacting laser beams. Also in the
case of Brillouin scattering, the nonlinear dispersion relation predicts a rather
weak interaction between the two beams, where the instability regions of the
two beams are more or less superimposed without dramatic differences in the growth rates.

In order to investigate the nonlinear dynamics of the interacting laser beams
in plasmas, we have carried out numerical simulations of the reduced system of 
equations (\ref{eq:nlse-coupled}) in two spatial dimensions, 
and have presented the results in Figs. 5--8. In these simulations, we
have used as an initial condition that either $A_1$ has a constant amplitude 
of $0.1$ and $A_2$ has a zero amplitude, or that both beams have a
constant amplitude of $0.1$ and that they initially have group velocities
at a right angle to each other. Due to symmetry reasons, it is sufficient
to simulate one vector component of $\vec{A}_j$, which we will denote $A_j$ ($j=1, \,2$).
The background plasma density is slightly
perturbed with a low-level noise (random numbers). We first consider 
stimulated Raman scattering, displayed in Figs. 5 and 6. The single beam 
case in Fig. 5 shows a growth of density waves mainly in the direction of
the beam, while a standing wave pattern is created in the amplitude of
the electromagnetic wave envelope, where maxima in the laser beam amplitude is 
(roughly) correlated with minima in the electron density. This is in line
with the standard Raman backscattering instability. The simulation is
ended when the plasma density fluctuations are large and self-nonlinearities
and kinetic effects are likely to become important. In Fig. 6, we show
the case with the two beams crossing each other at a right angle. In this case,
the wave pattern becomes slightly more complicated with local maxima
of the laser beam envelope amplitude correlated with local minima of the 
electron density. However, this pattern is very regular and there is
no clear sign of nonlinear structures in the numerical solution. 
We next turn to the case of stimulated Brillouin scattering, 
presented in Figs. 7 and 8. In this case, the waves grow
not only in the direction of the laser beam but also, with almost
the same growth rate, obliquely to the propagation direction of the laser beam. 
We see in the single beam case, presented in Fig. 7, that the envelope of the ion
beam becomes modulated in localized areas both in $y$ and $z$ directions,
and in the nonlinear phase at the end of the simulation, the laser
beam envelope has local maxima correlated with local minima of the
ion density. For the case of two crossed laser beams, displayed in
Fig. 8, we see a more irregular structure of the instability and that
at the final stage, local ``hot spots'' are created in which
large amplitude laser beam envelopes are correlated with local depletions
of the ion density.

\section{Summary}

In summary, we have investigated the instability and dynamics of two nonlinearly
interacting intense laser beams in an unmagnetized plasma. 
Our analytical and numerical results reveal that stimulated Raman 
forward and backward scattering instabilities are the dominating 
nonlinear processes that determine the stability of
intense laser beams in plasmas, where relativistic mass increases and 
the radiation pressure effects play a dominant role. 
Our nonlinear dispersion relation for two
interacting laser beams with different wavenumbers 
predicts a superposition of the instabilities for the single beams. 
The numerical simulation of the coupled nonlinear Schr\"odinger equations
for the laser beams and the governing equations for the slow plasma 
density perturbations in the presence of the radiation pressures, 
reveal that in the case of stimulated Raman scattering,
the nonlinear interaction between the two beams is weaker than for the
case of stimulated Brillouin scattering. The latter case lead to
local density cavities correlated with maxima in the electromagnetic wave
envelope.  The present results should be useful for understanding the nonlinear
propagation of two nonlinearly interacting laser beams in plasmas, as well as for the acceleration
of electrons by high gradient electrostatic fields that are created due
to stimulated Raman scattering instabilities in laser-plasma interactions.

\newpage

{\bf Figure captions}

\bigskip
\noindent

FIG. 1: The normalized (by $\omega_{p0}$) growth rates due to
stimulated Raman scattering (Case I) for single laser beams (upper panels) and
for two laser beams (lower panel), as a function of the 
wave numbers $K_y$ and $K_z$. The upper left and 
right panels show the growth rate for 
beam $\vec{A}_1$ and $\vec{A}_2$, respectively, where
the wave vector for $\vec{A}_1$ is $(k_y,\,k_z)=(6,\,0)\,\omega_{p0}/c$ and 
the one for $\vec{A}_2$ is $(k_y,\,k_z)=(0,\,4)\,\omega_{p0}/c$,
i.e. the two beams are launched in the $y$ and $z$ directions,
respectively. In the lower left panel, $\vec{A}_1$
and $\vec{A}_2$ are launched simultaneously at a 
perpendicular angle to each other, and in the lower right panel, the
two beams are counter-propagating. We used the normalized 
amplitudes $|\vec{A}_{10}|=|\vec{A}_{20}|=0.1$ and the electron thermal speed $v_{Te}=0.01c$.

\bigskip
\noindent

FIG. 2: The normalized (by $\omega_{p0}$) growth rates due to
stimulated Raman scattering (Case I) for single laser beams (upper panels) and
for two laser beams (lower panel), as a function of the 
wave numbers $K_y$ and $K_z$. The upper left and 
right panels show the growth rate for beam 
$\vec{A}_1$ and $\vec{A}_2$, respectively, where the 
wavenumber for $\vec{A}_1$ is $(k_y,\,k_z)=(5,\,0)\,\omega_{p0}/c$ and
the one for $\vec{A}_2$ is $(k_y,\,k_z)=(0,\,5)\,\omega_{p0}/c$.
In the lower left panel, two beams are launched at a 
perpendicular angle to each other, and in the lower right panel, the
two beams are counter-propagating. We used the normalized 
amplitudes $|\vec{A}_{10}|=|\vec{A}_{20}|=0.1$ and the electron thermal speed $v_{Te}=0.01c$.

\bigskip
\noindent

FIG. 3: The normalized (by $\omega_{p0}$) growth rates due to
stimulated Brillouin scattering (Case II) for single laser beams (upper panels) and
for two laser beams (lower panel), as a function of the 
wave numbers $K_y$ and $K_z$. The upper left and 
right panels show the growth rate for the beam $\vec{A}_1$ and $\vec{A}_2$,
respectively, where the wave number for $\vec{A}_1$ is $(k_y,\,k_z)=(6,\,0)\,\omega_{p0}/c$ 
and the one for $\vec{A}_2$ is $(k_y,\,k_z)=(0,\,4)\,\omega_{p0}/c$.
In the lower left panel, two beams are launched at a 
perpendicular angle to each other, and in the lower right panel, the
two beams are counter-propagating. We used the normalized 
amplitudes $|\vec{A}_{10}|=|\vec{A}_{20}|=0.1$, the ion to electron mass ratio $m_i/m_e=73440$ (Argon),
and the ion sound speed $c_s=3.4\times10^{-5}\,c$.

\bigskip
\noindent

FIG. 4: The normalized (by $\omega_{p0}$) growth rates due to
stimulated Brillouin scattering (Case II) for single laser beams (upper panels) and
for two laser beams (lower panel), as a function of the 
wave numbers $K_y$ and $K_z$. The upper left and 
right panels show the growth rate for beam $\vec{A}_1$ and $\vec{A}_2$, respectively,
where the wavenumber for $\vec{A}_1$ is $(k_y,\,k_z)=(5,\,0)\,\omega_{p0}/c$ and the one for $\vec{A}_2$
is $(k_y,\,k_z)=(0,\,5)\,\omega_{p0}/c$.
In the lower left panel, two beams are launched at a 
perpendicular angle to each other, and in the lower right panel, the
two beams are counter-propagating. We used the normalized 
amplitudes $|\vec{A}_{10}|=|\vec{A}_{20}|=0.1$, the ion to electron mass ratio $m_i/m_e=73440$ (Argon),
and the ion sound speed $c_s=3.4\times10^{-5}\,c$.

\bigskip
\noindent

FIG. 5: The amplitude of a single laser beam $|A_1|$ (left panels) and
the electron density $N_s$ (right panels) involving stimulated 
Raman scattering (Case I), at times 
$t=1.0\,\omega_{p0}^{-1}$, $t=30\,\omega_{p0}^{-1}$ and 
$t=60\,\omega_{p0}^{-1}$ (upper to lower panels). The laser beam
initially has the amplitude $A_1=0.1$ and wavenumber $(k_{1y},\,k_{1z})=(0,\,5)\,\omega_{p0}/c$. 
The electron density is initially perturbed with a small-amplitude noise
(random numbers) of order $10^{-4}$.

\bigskip
\noindent

FIG. 6: The amplitude of two crossed laser beams, 
$|A|=(|A_1|^2+|A_2|^2)^{1/2}$ (left panels) and
the electron density $N_s$ (right panels) involving stimulated 
Raman scattering (Case I), at times 
$t=1.0\,\omega_{p0}^{-1}$, $t=30\,\omega_{p0}^{-1}$ and 
$t=60\,\omega_{p0}^{-1}$ (upper to lower panels). The laser beams
initially have the amplitude $A_1=A_2=0.1$, and $A_1$ initially has the
wavenumber $(k_{1y},\,k_{1z})=(0,\,5)\,\omega_{p0}/c$ while $A_2$ has the 
wavenumber $(k_{2y},\,k_{2z})=(5,\,0)\,\omega_{p0}/c$. The electron density is initially 
perturbed with a small-amplitude noise (random numbers) of order $10^{-4}$.

\bigskip
\noindent

FIG. 7: The amplitude of a single laser beam $|A_1|$ (left panels) and
the electron density $N_s$ (right panels) involving stimulated 
Brillouin scattering (Case II), at times 
$t=1.5\,\omega_{p0}^{-1}$, $t=600\,\omega_{p0}^{-1}$ and 
$t=1200\,\omega_{p0}^{-1}$ (upper to lower panels). The laser beam
initially has the amplitude $A_1=0.1$ and wavenumber $(k_{1y},\,k_{1z})=(0,\,5)\,\omega_{p0}/c$. 
The ion density is initially perturbed with a small-amplitude noise
(random numbers) of order $10^{-4}$.

\bigskip
\noindent

FIG. 8: The amplitude of two crossed laser beams, 
$|A|=(|A_1|^2+|A_2|^2)^{1/2}$ (left panels) and
the electron density $N_s$ (right panels) involving stimulated 
Brillouin scattering (Case II), at times 
$t=1.0\,\omega_{p0}^{-1}$, $t=30\,\omega_{p0}^{-1}$ and 
$t=60\,\omega_{p0}^{-1}$ (upper to lower panels). The laser beams
initially have the amplitude $A_1=A_2=0.1$, and $A_1$ initially has the
wavenumber $(k_{1y},\,k_{1z})=(0,\,5)\,\omega_{p0}/c$ while $A_2$ has the 
wavenumber $(k_{2y},\,k_{2z})=(5,\,0)\,\omega_{p0}/c$. The electron density is initially 
perturbed with a small-amplitude noise (random numbers) of order $10^{-4}$.

\newpage

\begin{figure}[htb]
\centering
\includegraphics[width=10cm]{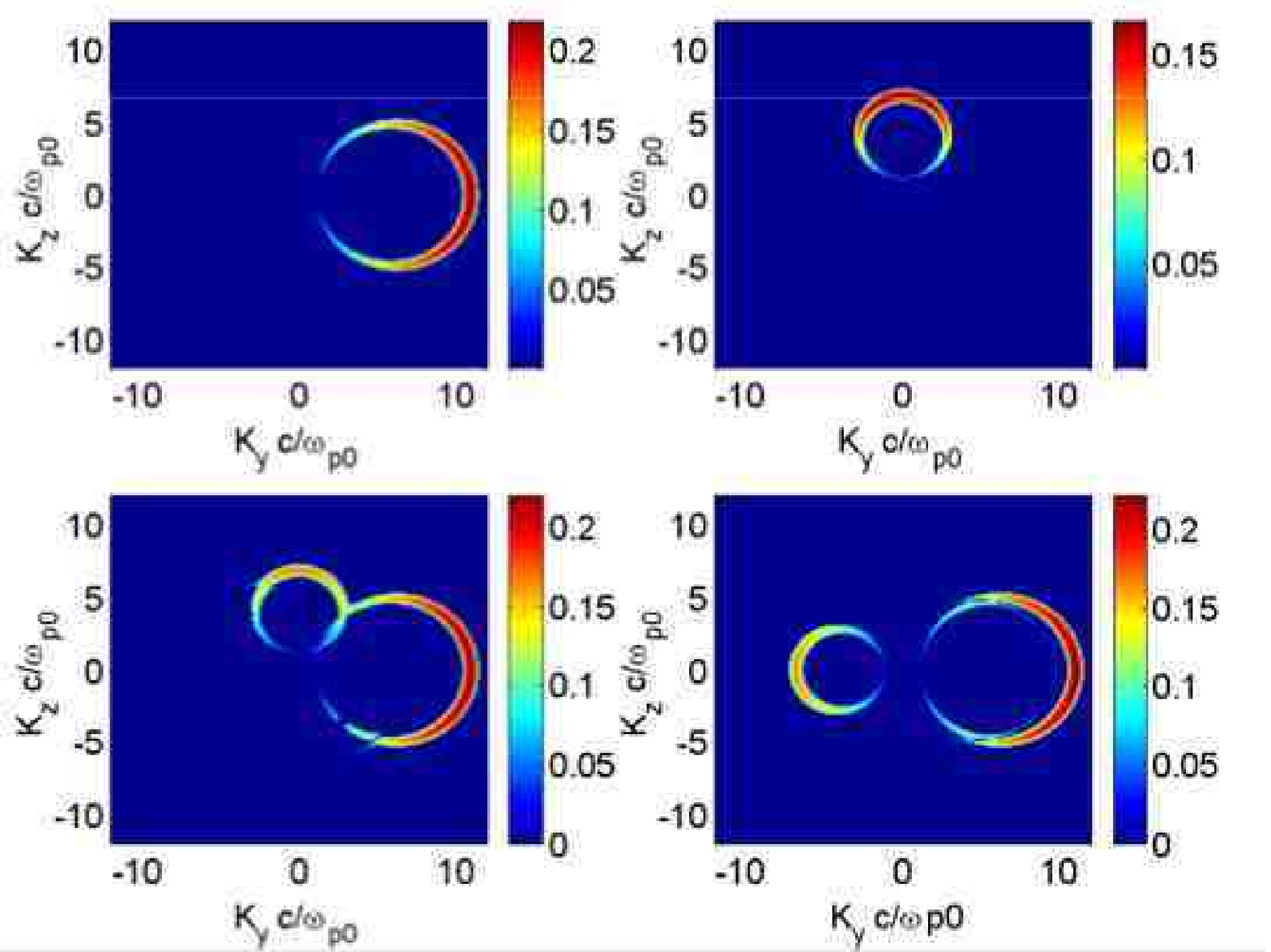}
\caption{}
\end{figure}

\begin{figure}[htb]
\centering
\includegraphics[width=10cm]{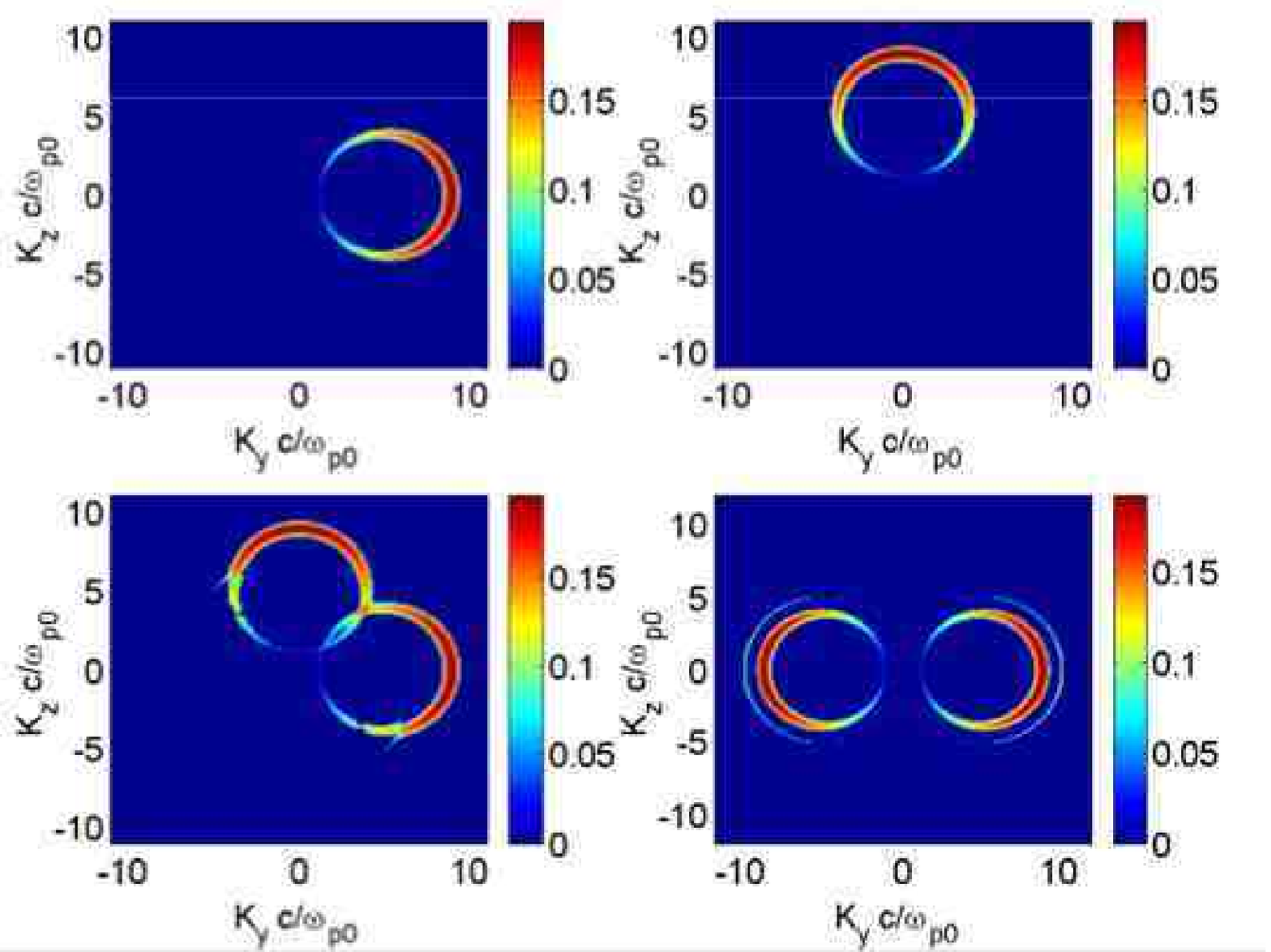}
\caption{}
\end{figure}

\begin{figure}[htb]
\centering
\includegraphics[width=10cm]{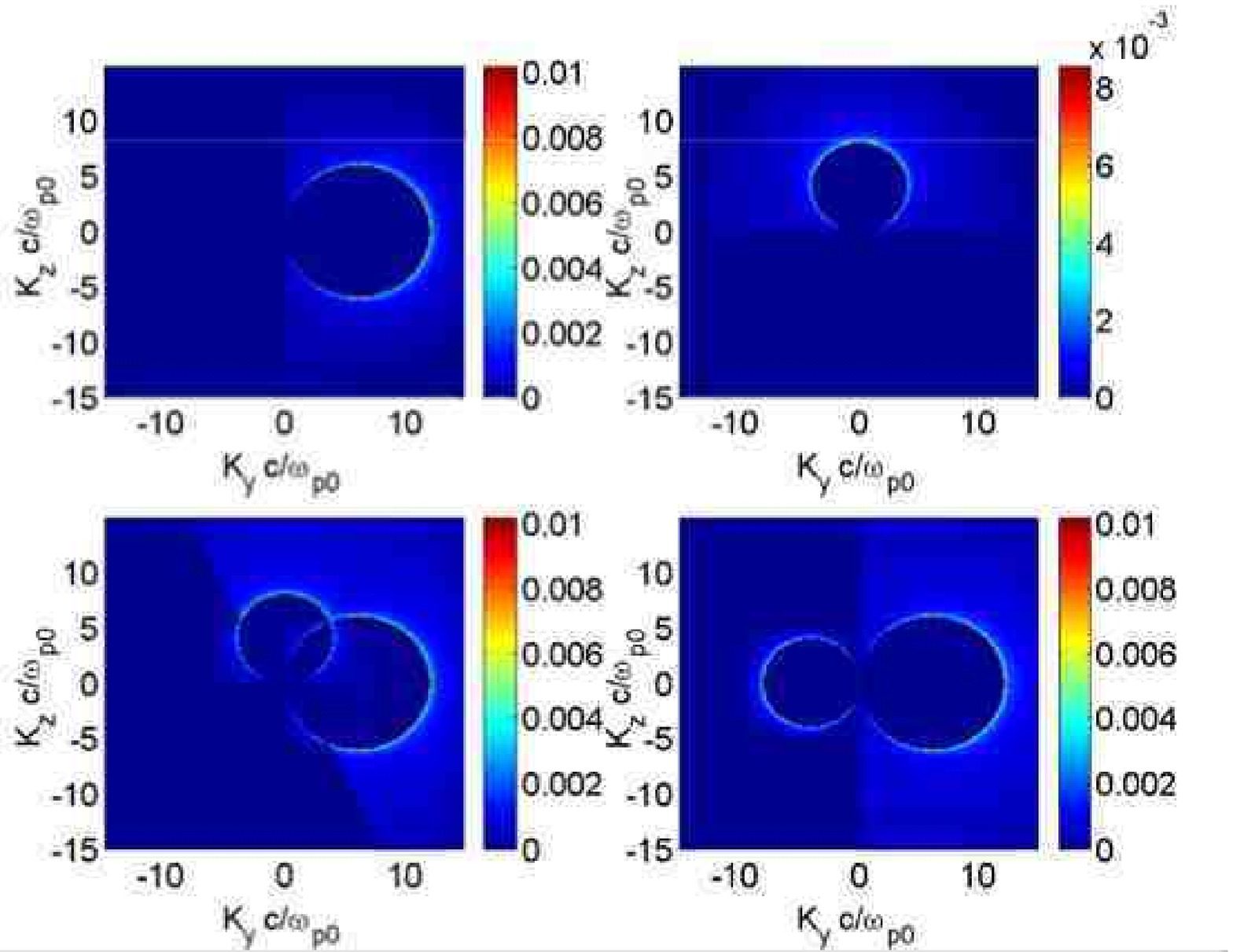}
\caption{}
\end{figure}

\begin{figure}[htb]
\centering
\includegraphics[width=10cm]{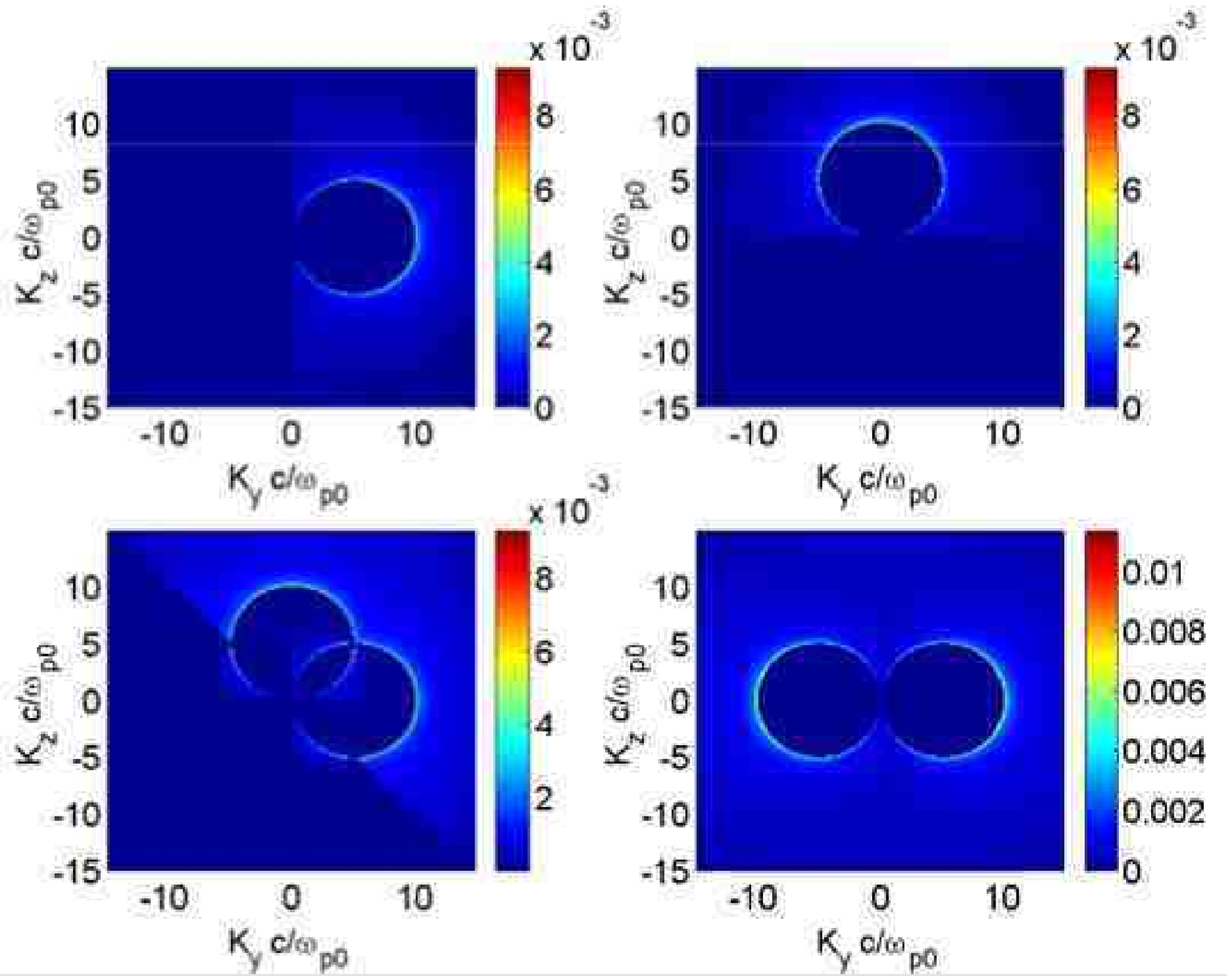}
\caption{}
\end{figure}

\begin{figure}[htb]
\centering
\includegraphics[width=10cm]{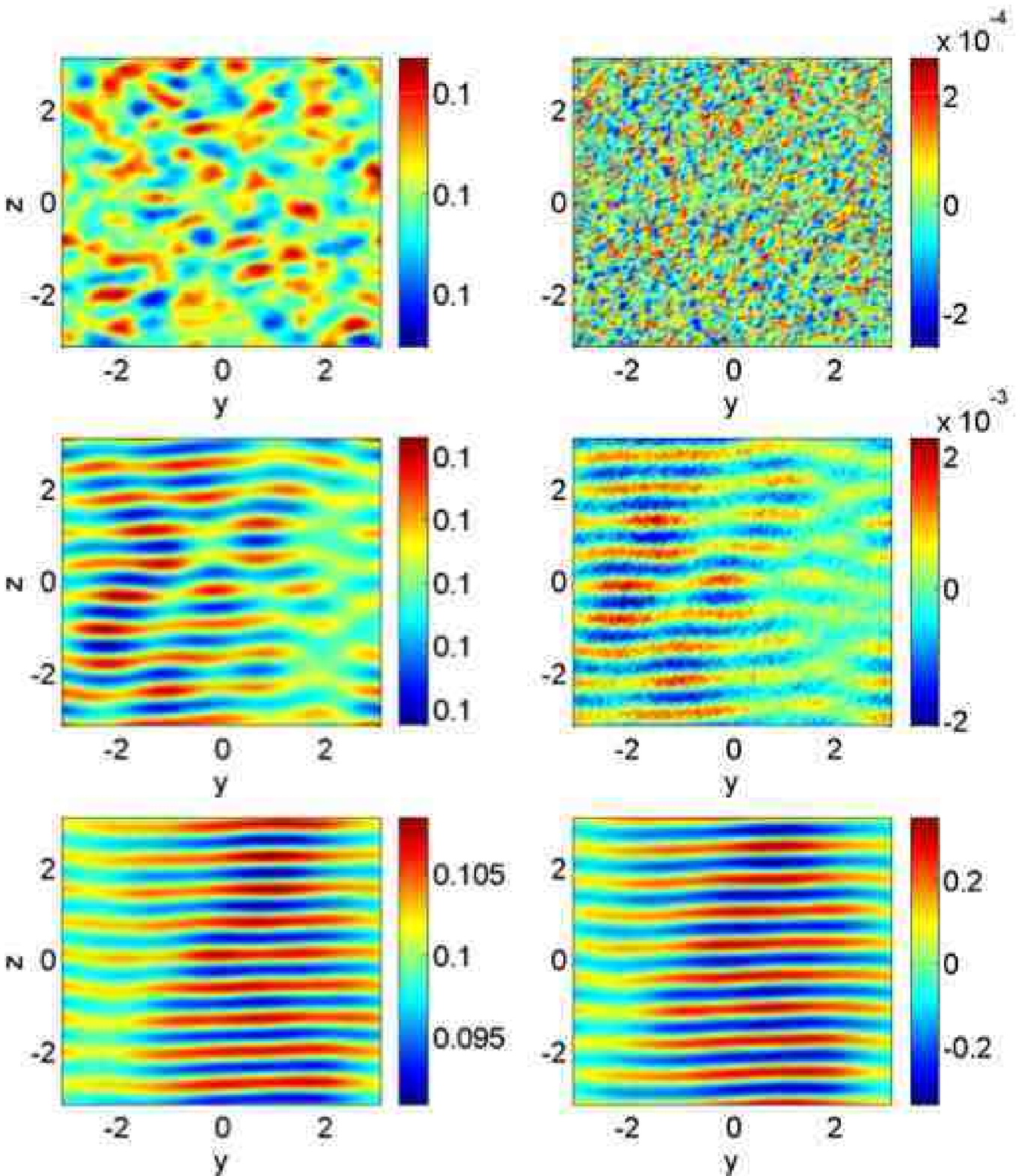}
\caption{}
\end{figure}

\begin{figure}[htb]
\centering
\includegraphics[width=10cm]{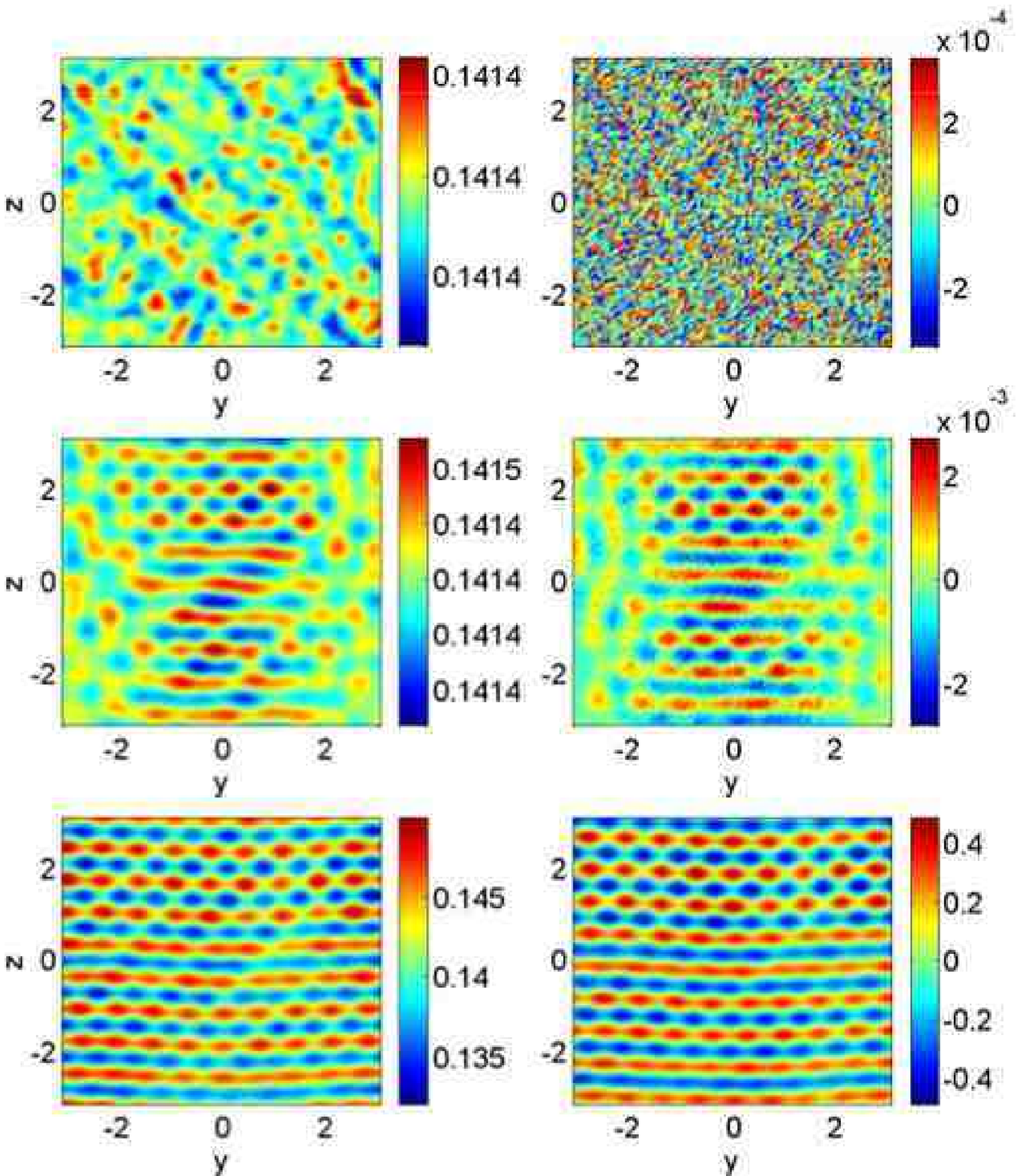}
\caption{}
\end{figure}

\begin{figure}[htb]
\centering
\includegraphics[width=10cm]{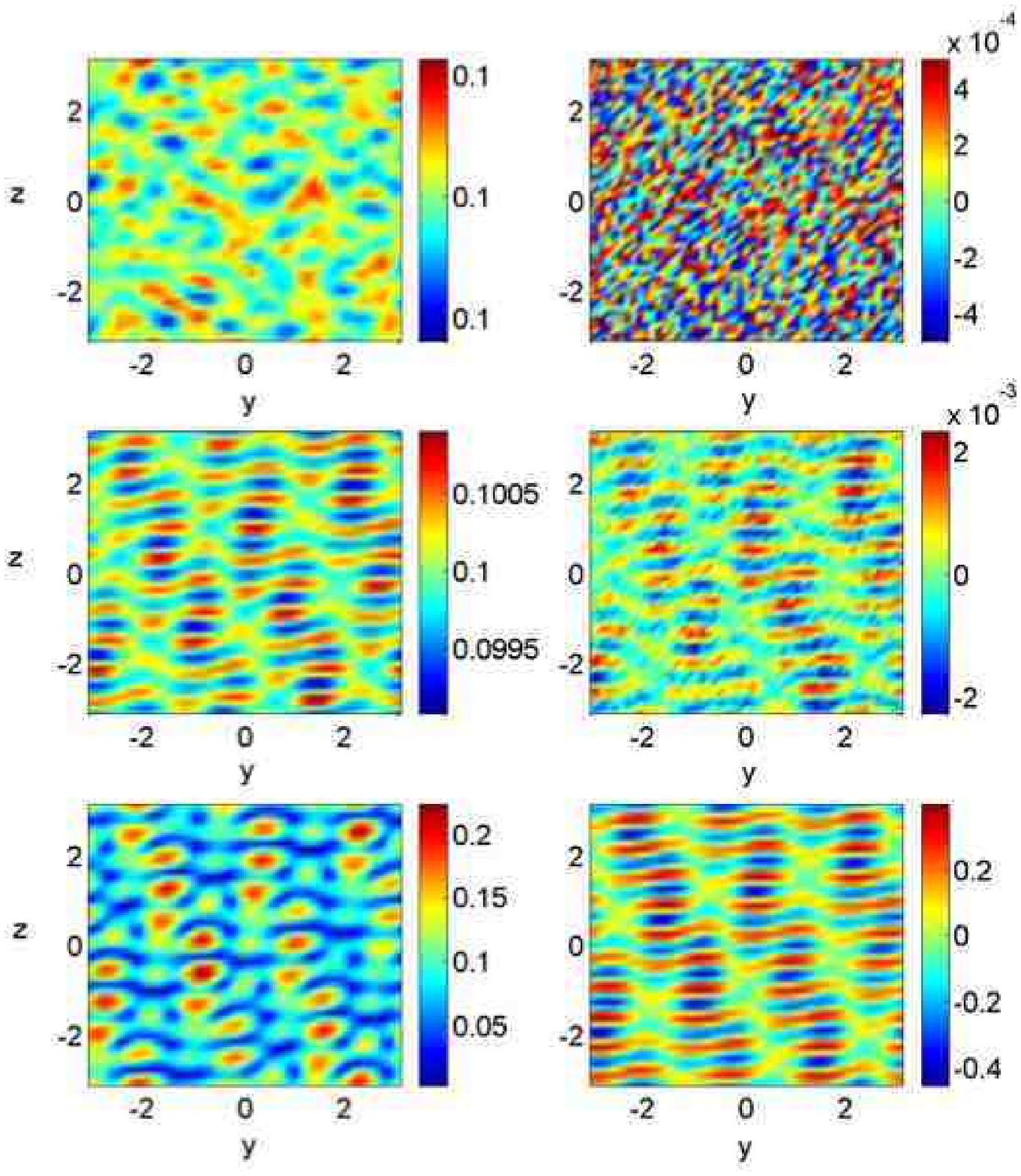}
\caption{}
\end{figure}

\begin{figure}[htb]
\centering
\includegraphics[width=10cm]{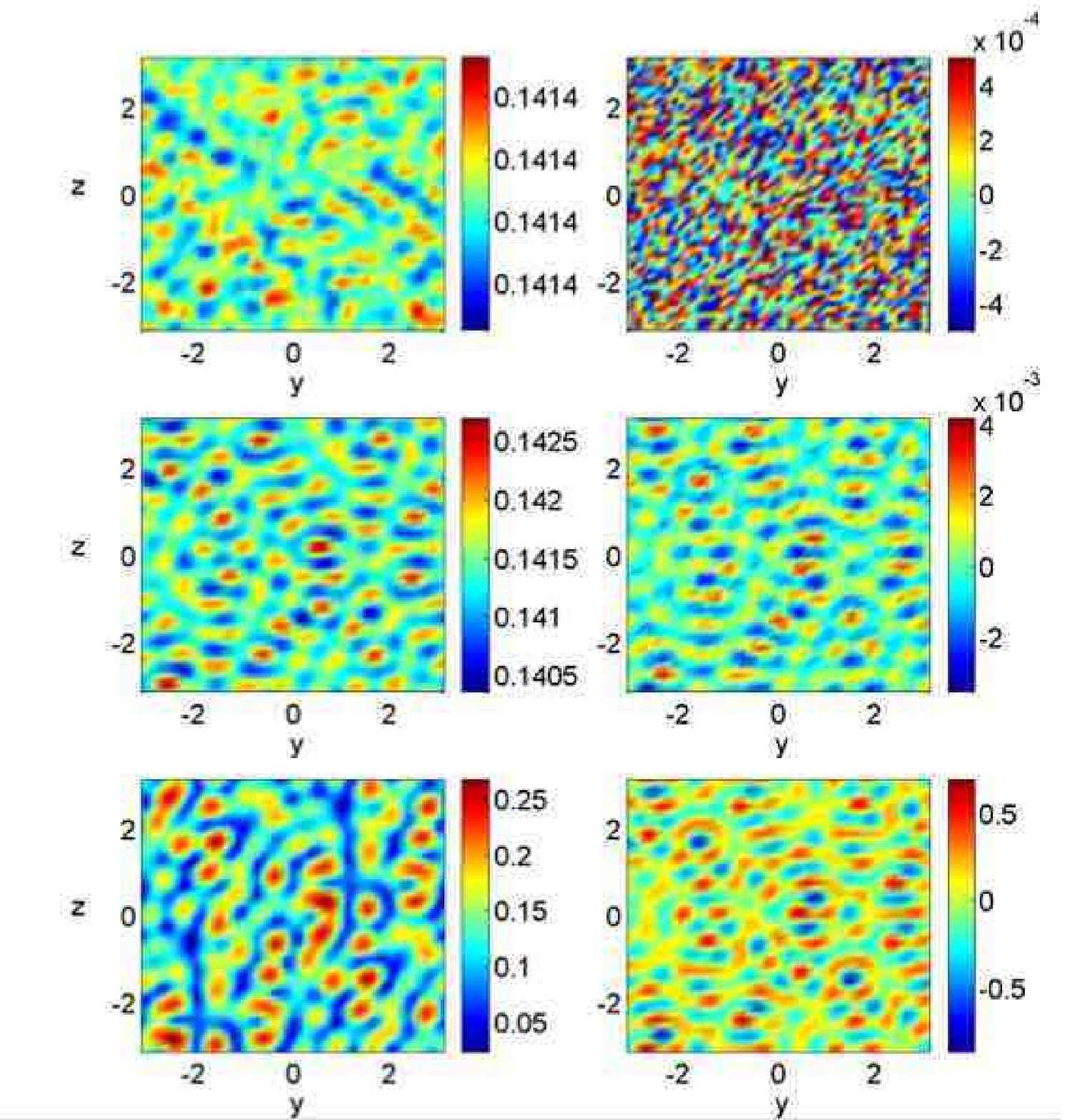}
\caption{}
\end{figure}

\end{document}